\documentclass[twocolumn,prl,amsmath,amssymb,nofootinbib]{revtex4}

\usepackage{dcolumn}
\usepackage{amsmath}
\usepackage{graphicx}
\usepackage{rotating}
\usepackage{amsfonts}
\usepackage{amssymb}
\usepackage[hang]{subfigure}

\usepackage{graphicx}
\usepackage{psfig,color}
\usepackage{epsfig}

\newcommand{\lwig}{\mbox{\,\raisebox{.3ex}
    {$<$}$\!\!\!\!\!$\raisebox{-.9ex}{$\sim$}\,}}
\newcommand{\gwig}{\mbox{\,\raisebox{.3ex}
    {$>$}$\!\!\!\!\!$\raisebox{-.9ex}{$\sim$}}\,}

\begin{document}

\preprint{}

\title {{\vskip -1cm\normalsize\rm\hfill DESY 06-098} 
\vskip 1cm Production and Detection of Axion-Like Particles at the VUV-FEL:\\[1ex] Letter of Intent} 

\author{Ulrich K\"otz}
\email{ulrich.koetz@desy.de}
\author{Andreas Ringwald\footnote{Corresponding author.}}
\email{andreas.ringwald@desy.de}
\author{Thomas Tschentscher}
\email{thomas.tschentscher@desy.de}
\affiliation{Deutsches Elektronen-Synchrotron DESY, Notkestra\ss e 85, D-22607 Hamburg, Germany}

\begin{abstract}
Recently, the PVLAS collaboration has reported evidence for an 
anomalously large rotation of the polarization 
of light generated in vacuum in the presence of a transverse magnetic field. 
This may be explained through the production of a 
new light spin-zero particle coupled to two photons. In this Letter of Intent, we propose to test 
this hypothesis by setting up a photon regeneration experiment
which exploits the photon beam of the Vacuum-UltraViolet Free-Electron Laser VUV-FEL, 
sent along the transverse magnetic field of a linear arrangement 
of dipole magnets of size $B\,L\approx 30$\,Tm.  
The high photon energies available at the VUV-FEL increase substantially the 
expected photon regeneration rate in the mass range implied by the PVLAS anomaly,  
in comparison to the rate expected at visible lasers of similar power.   
We find that the particle interpretation of the PVLAS result can be tested within a short running period.  
The pseudoscalar vs. scalar nature can be determined by varying the direction of the magnetic field 
with respect to
the laser polarization. The mass of the particle can be measured by running at different  
photon energies.  The proposed experiment offers a window of opportunity for a firm establishment or exclusion 
of the particle interpretation of the PVLAS anomaly before other experiments can compete.   
\end{abstract}


\maketitle

\section{Introduction and Motivation}

New very light spin-zero particles which are very weakly coupled to ordinary matter 
are predicted in many models beyond the Standard Model. Such light particles arise
if there is a global continuous symmetry in the theory that is spontaneously broken in the vacuum
--- a notable example being the axion~\cite{Weinberg:1978ma}, a pseudoscalar particle 
 arising from the breaking of a U(1) 
Peccei-Quinn symmetry~\cite{Peccei:1977hh}, 
introduced to explain the absence of $CP$ violation in strong interactions.
Such axion-like pseudoscalars couple to two photons via 
\begin{equation}
{\cal L}_{\phi \gamma \gamma} = - \frac{1}{4}\, g\, \phi\, F_{\mu \nu}\tilde{F}^{\mu \nu} = 
g\, \phi\, \vec{E}\cdot \vec{B} , 
\label{coupling}
\end{equation}
where $g$ is the coupling, $\phi$ is the field corresponding to the particle, 
$F_{\mu\nu}$ ($\tilde{F}^{\mu\nu}$) is the (dual) electromagnetic field strength tensor,
and $\vec{E}$ and $\vec{B}$ are the electric and magnetic fields, respectively. 
\begin{figure}[t]
\begin{center}
\includegraphics*[bbllx=86pt,bblly=637pt,bburx=298pt,%
bbury=707pt,width=8.cm,clip=]{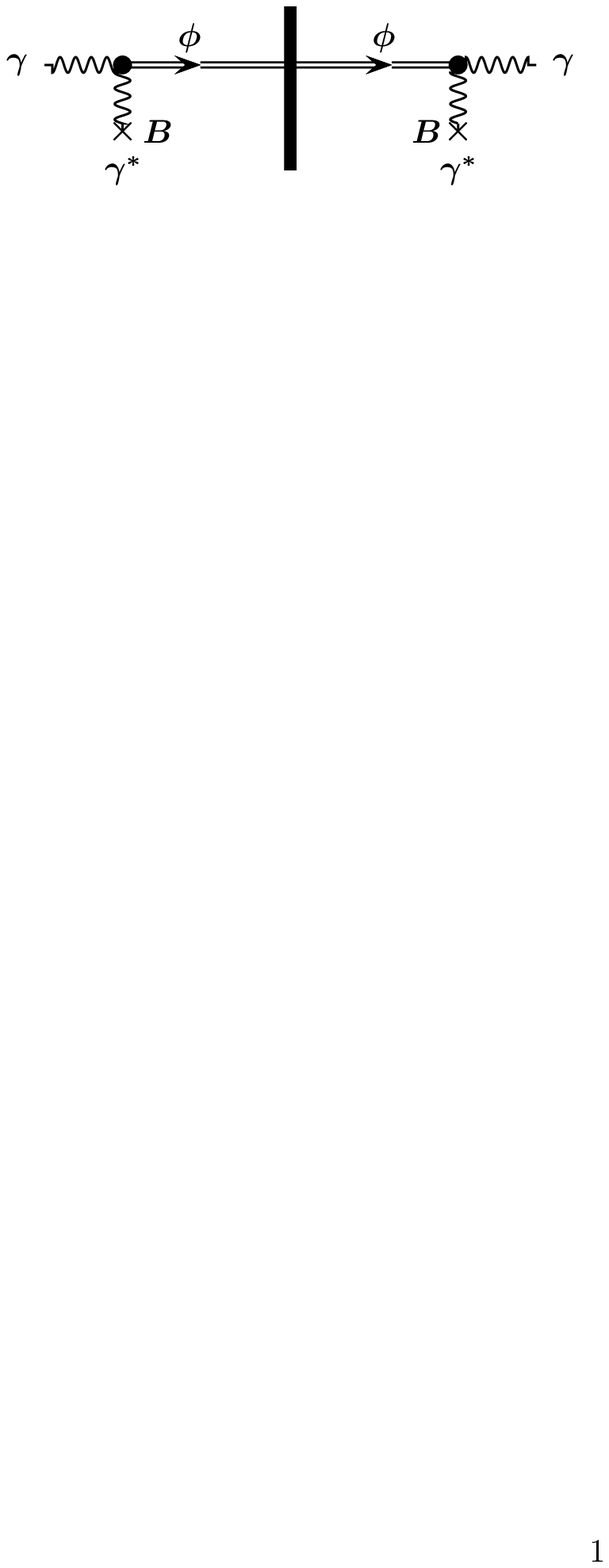}
\caption[...]{Schematic view of (pseudo-)scalar production through photon conversion 
in a magnetic field (left), subsequent travel through a wall, and final 
detection through photon regeneration (right). 
\hfill
\label{fig:ph_reg}}
\end{center}
\end{figure}
In the case of a scalar particle coupling to two photons, the interaction 
reads
\begin{equation}
{\cal L}_{\phi \gamma \gamma} =  \frac{1}{4}\, g\, \phi\, F_{\mu \nu} F^{\mu \nu} = 
g\, \phi\, \left( \vec{E}^2 - \vec{B}^2 \right).  
\label{coupling_scalar}
\end{equation}
Both effective interactions give rise to similar observable effects. In particular, 
in the presence of an external magnetic field, a photon of frequency $\omega$ may oscillate into a 
light spin-zero particle  of small mass $m_\phi < \omega$, and vice versa.  
The notable difference between a pseudoscalar and a scalar  is that it is
the component of the photon polarization parallel to the magnetic field that interacts in
the former case, whereas it is the perpendicular component in the latter case. 

\begin{figure}[t]
\begin{center}
\includegraphics*[bbllx=25,bblly=226,bburx=564,bbury=604,width=8.5cm]{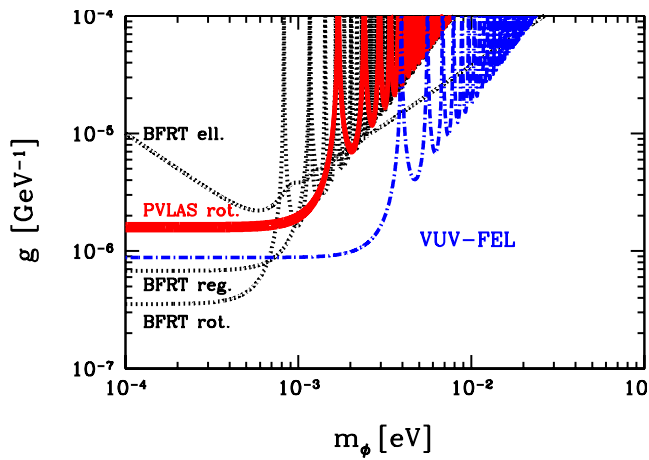}
\caption[...]{Two photon coupling $g$ of the (pseudo-)scalar versus its mass $m_\phi$. 
The upper limits from BFRT data~\cite{Cameron:1993mr}  
on polarization (rotation and ellipticity data; 95\,\% confidence level) 
and photon regeneration (95\,\% confidence level) are displayed as thick dots. 
The preferred values corresponding to the anomalous rotation signal observed by PVLAS~\cite{Zavattini:2005tm}  
are shown as a thick solid line. The projected 95\,\% confidence level upper limit which can be 
obtained with the  proposed experiment (see text)  is drawn as a dashed-dotted line. 
\hfill
\label{fig:ax_ph_lab_only}}
\end{center}
\end{figure}

The exploitation of this mechanism is the basic idea behind photon regeneration 
(sometimes called ``light shining through walls'') 
experiments~\cite{regeneration,VanBibber:1987rq}, see Fig.~\ref{fig:ph_reg}. Namely, if a beam of  
photons is shone across a magnetic field, a fraction of these photons will turn into (pseudo-)scalars.  
This (pseudo-)scalar beam can then propagate freely through a wall or another obstruction without being absorbed,  
and finally another magnetic field located on the other side of the wall can transform some of these (pseudo-)scalars 
into photons --- apparently regenerating these photons out of nothing. 
A pilot experiment of this type was carried out in Brookhaven using two prototype magnets for the 
Colliding Beam Accelerator~\cite{Ruoso:1992nx}. 
From the non-observation of photon regeneration, the Brookhaven-Fermilab-Rochester-Trieste
(BFRT) collaboration excluded values of the  coupling 
$g< 6.7  \times 10^{-7}\ {\rm GeV}^{-1}$, for $m_\phi \lwig 10^{-3}$ {\rm eV}~\cite{Cameron:1993mr} 
(cf. Fig.~\ref{fig:ax_ph_lab_only}), at the 90\,\% confidence level.   

\begin{figure}[t]
\begin{center}
\includegraphics*[bbllx=25,bblly=226,bburx=564,bbury=604,width=8.5cm]{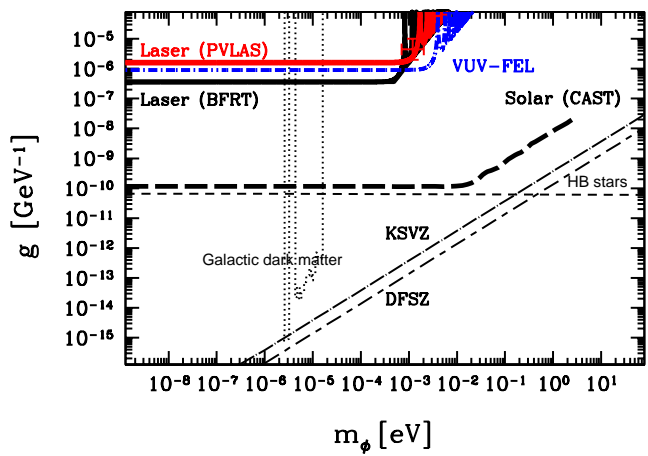}
\caption[...]{Exclusion region in mass $m_\phi$ vs.  coupling $g$ 
for various current and future experiments.
The laser experiments~\cite{Cameron:1993mr,Zavattini:2005tm} aim at  
(pseudo-)scalar production and detection in the laboratory.  
The galactic dark matter experiments~\cite{Eidelman:2004wy} exploit microwave cavities to detect 
pseudoscalars  under the assumption that these pseudoscalars are the dominant constituents of our galactic halo, 
and the solar experiments 
search for axions from the sun~\cite{Andriamonje:2004hi}. 
The constraint from horizontal branch (HB) stars~\cite{Raffelt:1985nk}  
arises from a consideration of stellar  
energy losses through (pseudo-)scalar production.
The predictions from two quite distinct QCD axion models, namely the KSVZ~\cite{Kim:1979if} 
(or hadronic) and the DFSZ~\cite{Zhitnitsky:1980tq} (or grand unified) one, 
are also shown.    
\hfill
\label{fig:ax_ph}}
\end{center}
\end{figure}

Recently, the PVLAS collaboration has reported an anomalous signal in 
measurements of the rotation of the polarization of photons in a magnetic field~\cite{Zavattini:2005tm}. 
A possible explanation of such an apparent vacuum magnetic dichroism is through the 
production of a light pseudoscalar or scalar, coupled to photons through Eq.~(\ref{coupling}) or 
Eq.~(\ref{coupling_scalar}), respectively.  Accordingly, 
photons polarized parallel (pseudoscalar) 
or perpendicular (scalar) 
to the magnetic field disappear, leading to a rotation of the 
polarization plane~\cite{Maiani:1986md}. The region quoted in Ref.~\cite{Zavattini:2005tm} that might explain 
the observed signal is  
\begin{eqnarray}
\label{PVLAS_coupling}
1.7 \times 10^{-6}\  {\rm GeV}^{-1}< & g & < 5.0 \times 10^{-6}\  {\rm GeV}^{-1},
\\[2ex]
\label{PVLAS_mass}
1.0 \times  10^{-3}\  {\rm eV} < & m_\phi & < 1.5 \times  10^{-3}\  {\rm eV},
\end{eqnarray}
obtained from a combination of
previous limits on $g$ vs. $m_\phi$ from a similar, but less sensitive polarization experiment 
performed by the BFRT collaboration~\cite{Cameron:1993mr} and the $g$ vs. $m_\phi$ curve 
corresponding to the PVLAS signal (cf. Fig.~\ref{fig:ax_ph_lab_only}).   

A particle with these properties presents a theoretical challenge. It is hardly compatible with a 
genuine QCD axion.  
Moreover, it must have very peculiar properties in order to evade the 
strong constraints on $g$ from stellar energy loss considerations~\cite{Raffelt:1985nk}
and from its non-observation 
in helioscopes such as the CERN Axion Solar Telescope~\cite{Andriamonje:2004hi,Raffelt:2005mt}  
(cf. Fig.~\ref{fig:ax_ph}). 
Its production in stars may be hindered, for example, 
if the $\phi\gamma\gamma$ vertex is suppressed at keV energies due to low scale 
compositeness of $\phi$, or if, in stellar interiors, 
$\phi$ acquires an effective mass larger than the typical photon energy, $\sim$~keV,  
or if the particles are trapped within stars~\cite{Masso:2005ym,Jain:2005nh,Jaeckel:2006id}.

Clearly, an independent and decisive experimental test of the pseudoscalar interpretation of
the PVLAS observation, without reference to axion production in stars 
(see~\cite{Dupays:2005xs,Kleban:2005rj}), is urgently needed. 
In Ref.~\cite{Rabadan:2005dm}, one of us (AR) was involved in the consideration of the possibility of 
exploiting powerful high-energy free-electron lasers (FEL) in a photon 
regeneration experiment\footnote{This idea has been considered first in Ref.~\cite{Ringwald:2001cp}.}  
to probe the region where the PVLAS signal could be explained in terms of the production of a light spin-zero 
particle.
In particular, it was emphasized that  the free-electron laser VUV-FEL~\cite{vuvfel} at DESY, 
which is designed to provide tunable radiation 
from the vacuum-ultraviolet (VUV; 10 eV) to soft X-rays (200 eV), will offer a unique and timely opportunity to probe the PVLAS result.   
Notably, the high photon energies available at the VUV-FEL increase substantially the 
expected photon regeneration rate in the mass range implied by the PVLAS anomaly,  
in comparison to the one expected at visible ($\sim 1$~eV) lasers.   
In this Letter of Intent,  we propose a corresponding photon regeneration experiment.

\section{Photon regeneration at the VUV-FEL}

The proposed experiment  is based on the assumption that the VUV-FEL can deliver photons 
with an energy $\omega = 38.7$~eV and an average photon flux ${\dot N}_0 = 6.5\times 10^{16}\,{\rm s}^{-1}$
(cf. Table~\ref{tab:vuvfel}).  
For the proposed photon regeneration experiment at the VUV-FEL, we 
study a linear arrangement of 12 normal conducting dipole magnets 
which are freely available at DESY\footnote{An option to use superconducting
magnets is under study.}. 
Each of these magnets has a  
magnetic field of $2.24$\,T and an integrated magnetic length of $1.029$\,m. 
The default arrangement consists of six plus six magnets, 
the beam absorber being placed between the first and second six. This 
arrangement corresponds to a magnetic field region of size $BL = 2B\ell= 27.66$\,Tm.
The proposed configuration is too large to fit into the VUV-FEL experimental hall. It has to be built  
on the ground before the entrance.  Correspondingly, the FEL beam line has 
to be extended to the proposed experiment.  

\begin{table}[t7]
\caption{\label{tab:vuvfel}
Achieved (2005) and expected (2007) VUV-FEL parameters.}
\begin{ruledtabular}
\begin{tabular}{lccc}
 &   & 2005 & 2007  \\
\hline
Bunch separation & [ns] & 1000 & 1000  \\
Bunches per train  & \# & 30 & 800  \\
Repetition rate & [1/s] & 5 & 10  \\
\hline
Photon wavelength & [nm] & 32 & 32  \\
Photon energy & [eV]   & 38.7 & 38.7  \\
Energy per pulse & [$\mu$J] & 10 & 50  \\
Photons per pulse & \# & $1.6\times 10^{12}$ & $8.1\times 10^{12}$  \\
Average flux & [1/s]  & $2.4\times 10^{14}$ & $6.5\times 10^{16}$  \\
\end{tabular}
\end{ruledtabular}
\end{table}

The photons leave the VUV-FEL with horizontal linear polarization.  In order to have a maximal 
coupling with a possible pseudoscalar/scalar, the magnetic field $\vec B$ of the magnets before the 
absorber should lie in the horizontal/perpendicular direction. We therefore foresee to exploit both 
possibilities of the magnetic field direction.

\begin{figure}[b]
\begin{center}
\includegraphics*[bbllx=47,bblly=233,bburx=578,bbury=607,width=8.5cm]{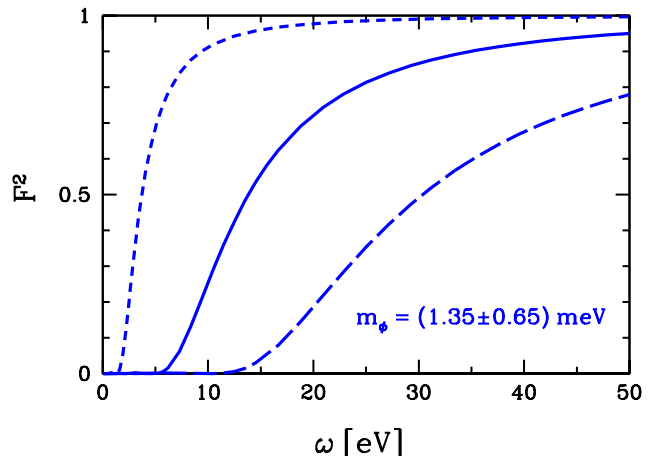}
\end{center}
\caption[...]{The second power of the form factor $F$, Eq.~(\ref{formfactor}), as a function of the 
laser frequency $\omega$, 
for fixed length $\ell =6$~m of the magnetic field region and different values of the pseudoscalar
mass $m_\phi$, corresponding to the central value, $m_\phi =1.35$~meV (solid), the lower value, 
$m_\phi=0.7$~meV (short dashed), and the upper value, $m_\phi=2.0$~meV (long dashed), 
of the range (\ref{PVLAS_mass}) suggested by PVLAS.
\hfill 
\label{form_factor}} 
\end{figure}

For the proposed experiment, the expected flux of regenerated photons is
\begin{eqnarray}
\nonumber 
\lefteqn{
{\dot N}_f \approx 1\times 10^{-4}\  {\rm s}^{-1}  \, F^2(q\ell )\, 
\left(\frac{{\dot N}_0}{6.5\times 10^{16}\ {\rm s}^{-1}}\right)
}
\\ \label{nf} 
&& \times 
\left( \frac{g}{10^{-6}\ {\rm GeV}^{-1}} \right)^4 
\left( \frac{B}{2.24\, {\rm T}}\right)^4  \left( \frac{\ell}{6\, {\rm m}}\right)^4 \, , 
\end{eqnarray}
where 
$q=m_\phi^2/(2\omega )$ ($\ll m_\phi$) is the momentum transfer to the magnet and
\begin{equation}
\label{formfactor}
F(q\ell ) = \left[\frac{\sin\left( \frac{1}{2}q\ell\right)}{\frac{1}{2}q\ell}\right]^2
\end{equation}
is a form factor which reduces to unity for small $q\ell$, 
corresponding to large $\omega$ (cf. Fig.~\ref{form_factor}) or small $m_\phi$, 
\begin{equation}
\label{small_mass}
m_\phi \ll \sqrt{\frac{2\,\pi\,\omega}{\ell}} = 3\times 10^{-3}\, {\rm eV}
    \sqrt{\left(\frac{ \omega}{38.7\, {\rm eV}}\right)  \left( \frac{6\, {\rm m}}{\ell }\right)  }
.
\end{equation}
For smaller $\omega$ or larger $m_\phi$, incoherence effects set in between the (pseudo-)scalar and the 
photon,  
the form factor getting much smaller than unity, 
severely reducing the regenerated photon flux~(\ref{nf}) (cf. Fig.~\ref{form_factor}). 
Therefore, a photon regeneration experiment exploiting the VUV-FEL beam has a unique advantage when compared 
to one using an ordinary laser operating near the visible ($\omega \sim 1$~eV): 
the sensitivity of the former extends to much larger 
masses\footnote{Running at conventional synchrotron radiation sources 
can extend the accessible mass range by more than an order of magnitude. However, they 
do not bring an advantage in terms of average photon flux. But  the continuous energy spectrum of 
the conventional synchrotron radiation reaching to high energies will result in a much larger background.}.  
In particular, the mass range~(\ref{PVLAS_mass}) implied by PVLAS is entirely 
covered, for $\omega =38.7$~eV photons (cf. Eq.~(\ref{small_mass})).  

In case a signal is found, one may use the possibility to tune the photon energy   
for a  determination of  the mass of the particle. 
This is done by lowering the energy of the FEL photons 
and observing the on-set of the incoherence expected around $\omega\approx m_\phi^2\ell/2\pi$. 
A determination of the form factor $F$ as a function of $\omega$ (cf. Fig.~\ref{form_factor}) will allow an 
extraction of $m_\phi$.

\section{Expected results}

The flux prediction~(\ref{nf}) uses the benchmark values for the VUV-FEL flux and
for the proposed magnetic field arrangement. For $g$ and $m_\phi$ in the parameter region preferred
by PVLAS,  Eqs.~(\ref{PVLAS_coupling}) and (\ref{PVLAS_mass}), this results 
in a rate of regenerated photons ranging from about 1 mHz up to 1 Hz.   

The very low predicted rates require therefore  a detector system with 
\begin{itemize}
\item a large single photon efficiency at $\sim 40$~eV, 
\item a short response time,  
\item and a low noise rate.
\end{itemize} 
Three detector options are being considered: electron multipliers, multi-channel plates, 
and avalanche photo diodes. One manufacturer quotes an efficiency of about 7\,\% for
electron multipliers. For the two other options, extrapolations point to a value of around 10\,\%.  
All three detectors show a response time in the 10 ns range.   
This short response time allows a  reduction of the noise rate by timing, 
exploiting the time structure of the photon beam.  
These detector performances, in particular the efficiencies and the response time, 
have to be studied at a beamline of the VUV-FEL as soon as  possible. 
At the same time, the general background rates in the VUV-FEL environment have to 
be studied as well. 

In the case of a non-observation of photon regeneration,  
for an assumed running time of $12\times 12$~h with an average photon flux of 
${\dot N}_0 = 6.5\times 10^{16}\,{\rm s}^{-1}$ at $\omega = 38.7$~eV, 
a 7\,\% single photon efficiency and zero background, 
the proposed experiment can establish a 95\% confidence 
limit of $g< 8.8\times 10^{-7}\ {\rm GeV}^{-1}$, for $m_\phi\,\lwig\, 3\times 10^{-3}$~eV.
In particular, the experiment is expected to be able to firmly 
exclude the particle interpretation of the PVLAS anomaly 
and to improve the current laboratory bound on $g$ in the $m_\phi\gwig 10^{-3}$~eV range  
(cf.~Fig.~\ref{fig:ax_ph_lab_only}).

\section{Conclusions}

The proposed experiment offers a window of opportunity 
for a firm establishment or exclusion of the particle interpretation of the PVLAS anomaly
in the near future. It takes essential advantage of unique properties of the VUV-FEL 
beam. The available VUV-FEL photon energies are just in the range where the 
photon regeneration rate is most sensitive to the hypothetical particle's mass. 
Moreover,  the well-defined beam of the VUV-FEL will not produce beam-related 
backgrounds. 

The experiment should be done soon, before other 
experiments~\cite{Cantatore:DESY,Baker:pc,BMV,LoI} can compete.  
A first step towards this goal is the study of possible detectors and their background rates.  
Finally, the proposed experiment could serve also as a test facility for an ambitious large scale photon regeneration 
experiment~\cite{Ringwald:2003ns}.

\vspace{0.15cm}
\centerline{\bf Acknowledgments}
\vspace{0.05cm}
We would like to thank Markus Ahlers, Josef Feldhaus, Ioannis Giomataris, Hermann Herzog, Rolf-Dieter Heuer,  
J\"org J\"ackel, Markus K\"orfer, Bernward Krause, Markus Kuster, Axel Lindner, Hinrich Meyer, 
Norbert Meyners, Dieter Notz, Alexander Petrov, 
Christian Regenfus, Horst Schulte-Schrepping, Albrecht Wagner, and Konstantin Zioutas  for valuable input.

\end{document}